\documentclass[12pt]{article}

\usepackage{amsmath,amsfonts,amssymb}

\begin{document}

\begin{titlepage}

	\begin{center}
		
		\vskip 0.4 cm
		
		\begin{center}
			{\Large  \bf Two Dimensional $F(\tilde{R})$ Ho\v{r}ava-Lifshitz Gravity}
		\end{center}
		
		\vskip 1cm
		
		\vspace{1em}  Josef Kluso\v{n}\footnote{Email address:
			klu@physics.muni.cz}\\
		\vspace{1em}\textit{Department of Theoretical Physics and
			Astrophysics, Faculty of Science,\\
			Masaryk University, Kotl\'a\v{r}sk\'a 2, 611 37, Brno, Czech Republic}
		
		\vskip 0.8cm
		
	\end{center}

	\begin{abstract}
		
		We study two-dimensional $F(\tilde{R})$ Ho\v{r}ava-Lifshitz gravity
		from the Hamiltonian point of view. We determine constraints structure
		with emphasis on the careful separation of the second class constraints
		and global first class constraints. We determine number of physical degrees of 
		freedom and also discuss gauge fixing of the global first class constraints.
		
	\end{abstract}
\end{titlepage}

\bigskip

\newpage

\def\mP{\mathcal{P}}
\def\bn{\mathbf{n}}
\newcommand{\bC}{\mathbf{C}}
\newcommand{\bD}{\mathbf{D}}
\def\tpi{\tilde{\pi}}
\def\hf{\hat{f}}
\def\tK{\tilde{K}}
\def\bmC{\bar{\mC}}
\def\tmG{\tilde{\mG}}
\def\tPi{\tilde{\Pi}}
\def\tmC{\tilde{\mC}}
\def\tPhi{\tilde{\Phi}}
\def\tv{\tilde{v}}
\def\mC{\mathcal{C}}
\def\bk{\mathbf{k}}
\def\tp{\tilde{p}}
\def\tr{\mathrm{tr}\, }
\def\tmH{\tilde{\mH}}
\def\tPsi{\tilde{\Psi}}
\def\tY{\mathcal{Y}}
\def\nn{\nonumber \\}
\def\bI{\mathbf{I}}
\def\tmV{\tilde{\mV}}
\def\e{\mathrm{e}}
\def\bE{\mathbf{E}}
\def\bX{\mathbf{X}}
\def\bH{\mathbf{H}}
\def\bY{\mathbf{Y}}
\def\bR{\bar{R}}
\def\hN{\hat{N}}
\def\tR{\tilde{R}}
\def\hK{\hat{K}}
\def\hnabla{\hat{\nabla}}
\def\hc{\hat{c}}
\def\mH{\mathcal{H}}
\def \Gi{\left(G^{-1}\right)}
\def\hZ{\hat{Z}}
\def\bz{\mathbf{z}}
\def\bK{\mathbf{K}}
\def\iD{\left(D^{-1}\right)}
\def\tmJ{\tilde{\mathcal{J}}}
\def\tr{\mathrm{Tr}}
\def\mJ{\mathcal{J}}
\def\tk{\tilde{k}}
\def\tGamma{\tilde{\Gamma}}
\def\partt{\partial_t}
\def\parts{\partial_\sigma}
\def\bG{\mathbf{G}}
\def\str{\mathrm{Str}}
\def\Pf{\mathrm{Pf}}
\def\bM{\mathbf{M}}
\def\tA{\tilde{A}}
\newcommand{\mW}{\mathcal{W}}
\def\bx{\mathbf{x}}
\def\by{\mathbf{y}}
\def \mD{\mathcal{D}}
\newcommand{\tZ}{\tilde{Z}}
\newcommand{\tW}{\tilde{W}}
\newcommand{\tmD}{\tilde{\mathcal{D}}}
\newcommand{\tN}{\tilde{N}}
\newcommand{\hC}{\hat{C}}
\newcommand{\hg}{g}
\newcommand{\hX}{\hat{X}}
\newcommand{\bQ}{\mathbf{Q}}
\newcommand{\hd}{\hat{d}}
\newcommand{\tX}{\tilde{X}}
\newcommand{\calg}{\mathcal{G}}
\newcommand{\calgi}{\left(\calg^{-1}\right)}
\newcommand{\hsigma}{\hat{\sigma}}
\newcommand{\hx}{\hat{x}}
\newcommand{\tchi}{\tilde{\chi}}
\newcommand{\mA}{\mathcal{A}}
\newcommand{\ha}{\hat{a}}
\newcommand{\tB}{\tilde{B}}
\newcommand{\hrho}{\hat{\rho}}
\newcommand{\hh}{\hat{h}}
\newcommand{\homega}{\hat{\omega}}
\newcommand{\mK}{\mathcal{K}}
\newcommand{\hmK}{\hat{\mK}}
\newcommand{\hA}{\hat{A}}
\newcommand{\mF}{\mathcal{F}}
\newcommand{\hmF}{\hat{\mF}}
\newcommand{\hQ}{\hat{Q}}
\newcommand{\mU}{\mathcal{U}}
\newcommand{\hPhi}{\hat{\Phi}}
\newcommand{\hPi}{\hat{\Pi}}
\newcommand{\hD}{\hat{D}}
\newcommand{\hb}{\hat{b}}
\def\I{\mathbf{I}}
\def\tW{\tilde{W}}
\newcommand{\tD}{\tilde{D}}
\newcommand{\mG}{\mathcal{G}}
\def\IT{\I_{\Phi,\Phi',T}}
\def \cit{\IT^{\dag}}
\newcommand{\hk}{\hat{k}}
\def \cdt{\overline{\tilde{D}T}}
\def \dt{\tilde{D}T}
\def\bra #1{\left<#1\right|}
\def\ket #1{\left|#1\right>}
\def\mV{\mathcal{V}}
\def\Xn #1{X^{(#1)}}
\newcommand{\Xni}[2] {X^{(#1)#2}}
\newcommand{\bAn}[1] {\mathbf{A}^{(#1)}}
\def \bAi{\left(\mathbf{A}^{-1}\right)}
\newcommand{\bAni}[1]
{\left(\mathbf{A}_{(#1)}^{-1}\right)}
\def \bA{\mathbf{A}}
\newcommand{\bT}{\mathbf{T}}
\def\bmR{\bar{\mR}}
\newcommand{\mL}{\mathcal{L}}
\newcommand{\mbQ}{\mathbf{Q}}
\def\mat{\tilde{\mathbf{a}}}
\def\mtF{\tilde{\mathcal{F}}}
\def \tZ{\tilde{Z}}
\def\mtC{\tilde{C}}
\def \tY{\tilde{Y}}
\def\pb #1{\left\{#1\right\}}
\newcommand{\E}[3]{E_{(#1)#2}^{ \quad #3}}
\newcommand{\p}[1]{p_{(#1)}}
\newcommand{\hEn}[3]{\hat{E}_{(#1)#2}^{ \quad #3}}
\def\mbPhi{\mathbf{\Phi}}
\def\tg{\tilde{g}}
\newcommand{\phys}{\mathrm{phys}}

\section{Introduction and Summary}
Study of two dimensional quantum gravity is very useful
when we can understand principles and puzzles of quantum gravity.
Two dimensional models are much simpler than four dimensional
gravity but share some interesting features with four dimensional gravity.
Further, two dimensional gravity plays a fundamental role
in the modern formulation of string theory
\cite{Polyakov:1981rd} where a propagating string in
$d-$dimensional flat target space-time can be described as a theory of $d-$
free scalar fields coupled to two dimensional gravity.

It is well known that there is no non-trivial gravitational dynamics
in space-time dimension lower than four. In three dimensions,
Riemann tensor is proportional to Ricci tensor and the source-free
theory is trivial. In two dimensions Einstein tensor is zero and
Einstein-Hilbert action is topological invariant.  As a result there
are no equations of motion and hence we cannot formulate meaningful
theory. In order to resolve this issue it was proposed in
\cite{Teitelboim:1983ux} that the appropriate model for two
dimensional gravity is the constant curvature equation
${}^{(2)}R-2\Lambda=0$, where ${}^{(2)}R$ denotes the two
dimensional Ricci scalar. In order to study quantum properties of
this theory we need an action principle from which this equation
can be derived. It turned out that the only invariant action is the
non-geometric action that involves scalar field $\Phi$ as a Lagrange
multiplier
\begin{equation}
S=\int d^2x \Phi({}^{(2)}R-2\Lambda) \ ,
\end{equation}
that leads to desired equations of motion when we perform variation
with respect to $\Lambda$.  The exact solution of this model was
found in \cite{Henneaux:1985nw}.

Few years ago P. Ho\v{r}ava formulated its famous model of power
counting renormalizable theory of gravity known as
Ho\v{r}ava-Lifshitz gravity (HL) \cite{Horava:2009uw} which is the theory
of gravity that is not invariant under full four dimensional
diffeomorphism but under reduced group of diffeomorphism known as a
foliation preserving diffeomorphism in order to have theory with
anisotropic scale invariance. In fact, the requirement of the
anisotropic scale invariance is central for the power counting
renormalizability of this theory. On the other hand the reduced
group of diffemorphism has very strong impact on the structure of
the theory since there are additional modes with important
phenomenological and theoretical consequences on the consistency of
the theory.

 This theory has an
improved behavior at high energies due to the presence of the higher
order spatial derivatives in the action which implies that the
theory is not invariant under full diffeomorphism but it is
invariant under so called foliation preserving diffeomorphism
($\mathrm{Diff}_\mF$)
\begin{equation}\label{DiffF}
t'=f(t) \ , \quad x'^i=x^i(\bx,t) \ .
\end{equation}
This property offers the possibility that the space and time
coordinates have different scaling at high energies
\begin{equation}
t'=k^{-z}t \ , \quad x'^i=k^{-1}x^i \ ,
\end{equation}
where $k$ is a constant. Consequence of this fact is that in $3+1$
dimensions the theory contains terms with $2$ time derivatives and
at least $2z$ spatial derivatives since the minimal amount of the
scaling anisotropy that is needed for the power-counting
renormalizability of this theory is $z=3$.
 Then collecting all
terms that are invariant under $\mathrm{Diff}_\mF$ symmetry leads to
the general action \cite{Blas:2009qj,Blas:2010hb}
\begin{eqnarray}
S=\frac{M_p^2}{2} \int dt d^3\bx N \sqrt{g}K_{ij}
\mG^{ijkl}K_{kl}-S_V  \ ,
\end{eqnarray}
where
\begin{equation}
K_{ij}=\frac{1}{2N}(\partial_t g_{ij}-D_iN_j-D_jN_i) \ ,
\end{equation}
and where we introduced generalized De Witt metric $\mG^{ijkl}$
defined as \cite{Horava:2008ih}
\begin{equation}
\mG^{ijkl}=\frac{1}{2}(g^{ik}g^{jl}+g^{il}g^{jk})-\lambda
g^{ij}g^{kl} \ ,
\end{equation}
where $\lambda$ is an arbitrary real constant. Finally note that
$D_i$ is  the covariant derivative defined with the help of the
metric $g_{ij}$.

 The action $S_V$ is the potential term action in
the form
\begin{equation}
S_V=\frac{M_p^2}{2}\int dt d^3\bx N\sqrt{g}\mV=\frac{M_p^2}{2}\int
dt d^3\bx N\sqrt{g}\left(\mL_1+\frac{1}{M^2_*}\mL_2+
\frac{1}{M^4_*}\mL_3\right) \ ,
\end{equation}
where $\mL_n$ contain all terms that are invariant under foliation
preserving diffeomorphism and where $\mL_n$ contain $2n$ derivatives
of the ADM variables $(N,g_{ij})$. In the UV when $k\gg M_*$ the
dominant contributions come from the higher derivative terms that
lead to the modified dispersion relation $\omega^2\propto k^6$  that
implies that this theory is power counting renormalizable. In the
opposite regime $k\ll M_*$ the dispersion relation is relativistic
and it can be shown that the theory have regions in the parameter
space where  it is in agreement with observation.

This theory has very interesting property which is the presence of
the vector $a_i$ that contains spatial derivative of lapse $N$.
These terms are forbidden in the theory invariant under full
diffeomorphism which implies an existence of the local first class Hamiltonian
constraint. In case of HL gravity the canonical structure is much
more complicated as was shown previously in 
\cite{Chaichian:2015asa,Kluson:2010nf,Mukohyama:2015gia,Donnelly:2011df}.
More precisely, two second class constraints were identified which 
should be solved for lapse $N$ and conjugate momentum. However generally 
this constraint is  second order 
partial differential equation for lapse  whose explicit solution was very difficult to find. For that reason it is  instructive to perform an analysis of
much simpler models as is for example two dimensional HL gravity.
This was done previously in \cite{Li:2015itk}. Our goal is to
generalize this analysis to the case of two dimensional $f(\tR)-$ HL
gravity which is more complex and allows local degrees of freedom on the
reduced phase space.
We also discuss the subtle point of the global first class
constraints \cite{Donnelly:2011df}. We argue that in order to solve the second
class constraints we have to fix these global constraints. This is 
very important observation for the structure of the reduced phase space  when we 
determine equations of motion for variables that define reduced phase
space and we show that it takes rather complicated form. As a result 
 we are not able to derive Hamiltonian on the reduced phase space that is apparently 
non-local due to the necessity to fix global first class constraints with 
global gauge fixing functions.
 
As the check of the validity of our procedure we discuss two special cases of the choice of the parameters in this theory. The first one corresponds
to the diffeomorphism invariant two dimensional $f(R)$ theory. We determine the canonical structure of this theory and we argue that it has the same form as in seminal papers \cite{Teitelboim:1983ux,Henneaux:1985nw}. Then we proceed to the analysis of the reduced phase space theory when we fix 
all first class constraints. We show that there are no physical degrees of freedom on the reduced phase space and we show that with suitable chosen gauge fixing function we derive equations for lapse and for scalar field that are in agreement with the equations derived in 
\cite{Almheiri:2014cka} which is also nice consistency  check of our analysis. 
Finally we consider the case when the function that defines $f(\tR)$ theory is identically equal to one. This situation corresponds to the non-projectable HL gravity in two dimensions that was analyzed previously in 
\cite{Li:2015itk}. We perform the canonical analysis of this theory from different point of view with emphasis on the existence of two global first class constraints and their gauge fixing. Solving all constraints we show that there are no physical degrees of freedom left and that these constraints lead to the solution that is in agreement with the analysis 
performed in \cite{Li:2015itk}.

Let us outline our results. We performed canonical analysis of two dimensional $f(\tilde{R})$ HL gravity and we show that the equations on the reduced phase space are rather complicated and contain integration over the whole space interval as a consequence of the gauge fixing of the global constraints. We mean that this is very important result that should be valid in higher dimensional non-projectable theory as well and which certainly makes the canonical analysis even more complicated than it is. 

This paper is organized as follows. In the next section (\ref{second}) we 
introduce two dimensional $f(\tR)$ HL gravity and define basics notations. Then in section (\ref{third}) we perform Hamiltonian analysis of this theory and determine all constraints. In section (\ref{fourth}) we 
consider special values of parameters that correspond to $f(R)-$gravity in two dimensions and we perform its Hamiltonian analysis. Finally in section 
(\ref{fifth}) we analyze pure non-projectable HL gravity in two dimensions
from Hamiltonian point of view. 
\section{Two Dimensional $F(\tR)-$Horava-Lifshitz Gravity}
\label{second}
In this  section we formulate two dimensional HL $f(\tR)$ gravity.
 Clearly the 
action for this system is the special case of higher dimensional
$f(\tR)$ HL gravities that were studied before, see for example
\cite{Kluson:2009rk,Kluson:2009xx,Kluson:2010xx,Elizalde:2010ep}.
Let us consider following model of two dimensional non-projectable HL
$f(\tR)$ gravity
\begin{equation}
S=\frac{1}{\kappa}\int dt dx N\sqrt{g}f(\tR) \ , 
\end{equation}
where $\kappa=8\pi G_N$ and where $\tR$ is defined as
\begin{equation}
\tR=\mL_K-\mL_V \ , 
\end{equation}
where
\begin{equation}
\mL_K=K_{ij}K^{ij}-\lambda
K^2+\frac{2\mu}{\sqrt{g}N}\partial_\mu(\sqrt{g}N n^\mu
K)-\frac{2\mu}{\sqrt{g}N}\partial_i (\sqrt{g}g^{ij}\partial_j N) \ ,
\end{equation}
with $K_{ij}=\frac{1}{2N}(\partial_t g_{ij}-D_i N_j-D_j N_i)$ where
$D_i$ denotes the covariant derivative of the metric $g_{ij}$ and
$N^i$ is the shift vector $N^i=g^{ij}N_j$. Finally $n^\mu$ is future
pointing normal vector  to the surface $\Sigma_t$ that in ADM
variables is equal to $n^0=\frac{1}{N} \ , n^i=-\frac{N^i}{N}$.
Finally $\mu$ is a free parameter that approaches $1$ in the low
energy limit.

Let us now discuss the potential term $\mL_V$ that is made of $R$,
$D_i$ and $a_i=\frac{\partial_i N}{N}$ where $R$ is Ricci scalar of
the leaves $t=\mathrm{const}$ that identically vanishes at one
dimension $R=0$. It can be shown \cite{Li:2015itk}
that  in $d=1$
dimensions $\mL_V$ has the form
\begin{equation}
\mL_V=2\Lambda-\beta a_i a^i \ ,
\end{equation}
where $\Lambda$ is cosmological constant and $\beta$ is another
dimensionless coupling constant.

To deal with $f(\tR)$ gravity in two dimensions we introduce two
scalar fields and write the action as
\begin{eqnarray}
S&=&\frac{1}{\kappa}\int dt dx N \sqrt{g} (f(A)+B(\tR-A))= \nonumber
\\
&=&\frac{1}{\kappa}\int dt dx N\sqrt{g} (f(A)-BA+
B(K_{ij}K^{ij}-\lambda K^2)-2\mu \partial_\mu Bn^\mu K+2\mu
\partial_i Bg^{ij} a_j- \nonumber
\\
&-&2\Lambda B+\beta B a_i a^i) \ . \nonumber \\
\end{eqnarray}
In $1+1$ dimensions $g_{ij}$ has only one components that we denote,
following \cite{Li:2015itk} as
\begin{equation}
\gamma\equiv \sqrt{g_{11}} \ , \quad g_{11}=\gamma^2 \ , \quad
g^{11}=\frac{1}{\gamma^2}
\end{equation}
so that we have following non-zero component of $\Gamma^1_{11}$
\begin{equation}
\Gamma^1_{11}=\frac{1}{2}g^{11}\partial_1 g_{11}=\frac{1}{\gamma}
\gamma' \ ,  \quad \gamma'\equiv \frac{\partial}{\partial x}\gamma \ . 
\end{equation}
Then we easily find that the action has the form 
\begin{eqnarray}\label{Sact1}
S=\frac{1}{\kappa}\int dt dx N\gamma (f(A)-BA+ B(1-\lambda)K^2
-2\mu \nabla_n BK+2\mu\frac{1}{\gamma^2}B'a-2\Lambda B+\beta Ba^2
\frac{1}{\gamma^2}) \ , \nonumber \\
\end{eqnarray}
where 
\begin{equation}
K=g^{11}K_{11}=
\frac{1}{N}(\frac{\dot{\gamma}}{\gamma}-\frac{N'_1}{\gamma^2}+
\frac{\gamma'}{\gamma^3}N_1) \ , \quad 
\nabla_nB=\frac{1}{N}(\dot{B}-N^1 B' ), \quad a\equiv
a_1 \ ,
\end{equation}
where $\dot{B}=\partial_t B \ , B'=\partial_1 B$.
The action (\ref{Sact1}) will be starting point of our canonical analysis
that will be performed in the next section. 
\section{Hamiltonian Analysis}\label{third}
Now we proceed to the Hamiltonian analysis of the theory specified
by the action (\ref{Sact1}). Before we do it
it is useful to simplify this action with the help of the fact that the variable $A$  has no dynamics and can be eliminated by
solving its equation of motion. In more details, the equation of motion for $A$ has the form 
\begin{equation}\label{eqA}
\frac{df}{dA}-B=0 \ . 
\end{equation}
If we presume that there is a function $\Psi$ that is inverse 
to $\frac{df}{dA}$ we find that  the equation (\ref{eqA})  has the solution 
\begin{equation}
A=\Psi(B) \ .
\end{equation}
Inserting this solution into  the action (\ref{Sact1}) 
we obtain the final form of the action
\begin{eqnarray}\label{actHam}
S=\frac{1}{\kappa}\int dt dx N\gamma \left( B(1-\lambda)K^2 -2\mu
\nabla_n BK+2\mu\frac{1}{\gamma^2}B'a-U(B)+\beta Ba^2
\frac{1}{\gamma^2}\right) \ , \nonumber \\
\end{eqnarray}
where 
\begin{equation}
U(B)=f(\Psi(B))-B\Psi(B) \ . 
\end{equation}
Starting with the action (\ref{actHam}) we find following conjugate
momenta
\begin{eqnarray}
\pi_N&=&\frac{\delta L}{\delta \dot{N}}\approx 0 \ , \quad 
\pi^1=\frac{\delta L}{\delta \dot{N}_1}\approx 0 \ , \nonumber \\
\pi&=&\frac{\delta L}{\delta
\dot{\gamma}}=\frac{2B}{\kappa}(1-\lambda)K-\frac{2\mu}{\kappa}
\nabla_n B \ ,  \quad P=\frac{\delta L}{\delta
\dot{B}}=-\frac{2\mu}{\kappa}\gamma K \ .
 \nonumber \\
\end{eqnarray}
Then it is easy to perform Legendre transformation in order to find
corresponding Hamiltonian
\begin{eqnarray}
H=\int dx(\pi\dot{\gamma}+P\dot{B}-\mL)=\int dx (N\mH_T+N_1\frac{1}{\gamma^2}\mH_1) \ , 
\nonumber \\
\end{eqnarray}
where 
\begin{eqnarray}
\mH_T&=&-\frac{\kappa}{4\mu^2\gamma}B(1-\lambda)P^2-\frac{\kappa}{2\mu}P\pi
-\frac{2\mu}{\kappa\gamma}B'a
+\frac{\gamma}{\kappa}U(B)-\frac{\beta}{
\kappa}\frac{Ba^2}{\gamma} \ ,  \nonumber \\
\mH_1&=&-\gamma\pi'+P B' \ 
\nonumber \\
\end{eqnarray}
using
\begin{eqnarray}
K=-\frac{\kappa}{2\mu\gamma }P \ , -\frac{2\mu}{\kappa} \nabla_n
B=\pi+\frac{2B}{2\mu\gamma}(1-\lambda)P \ . \nonumber \\
\end{eqnarray}
Now we have to analyze the requirement of the preservation of the
primary constraints $\pi_N\approx 0 \ , \pi^1\approx 0$
\begin{eqnarray}
\partial_t\pi_N&=&\pb{\pi_N,H}\nonumber \\
&=&-\mH_T-\frac{2\mu}{\kappa
\gamma}\frac{N'}{N}B'-\left(\frac{2\mu}{\kappa}\frac{B'}{\gamma}\right)'
-\frac{2\beta}{\kappa}\frac{B}{\gamma}a^2-\left(
\frac{2\beta}{\kappa}\frac{B}{\gamma}a\right)' \equiv -\mC\approx 0 \ , 
 \nonumber \\
 \partial_t \pi^i&=&\pb{\pi^1,H}=-\mH_1\approx 0 \ . \nonumber \\
\end{eqnarray}
Note that $\mC$ obeys an important relation
\begin{equation}
\int dx N \mC=\int dx N\mH_T \ 
\end{equation}
using integration by parts and also the fact that we presume
suitable asymptotic behavior of all fields so that the contributions from 
spatial infinities can be ignored. 
As in higher dimensional non-projectable HL gravity we introduce the
global primary constraint
\begin{equation}
\Pi_N=\int dx \pi_N N
\end{equation}
and split original constraint $\pi_N$ into $\infty-1$ local ones
\begin{equation}
\tpi_N=\pi_N-\frac{\gamma}{\int dx \gamma N}\Pi_N
\end{equation}
that obeys the relation
\begin{equation}
\int dx N\tpi_N=0 \ .
\end{equation}
Then the requirement of the preservation of the primary constraint
$\Pi_N$ implies
\begin{equation}
\partial_t \Pi_N=\pb{\Pi_N,H}=-\int dx N\mH_T\equiv-\Pi_T\approx 0
\end{equation}
using
\begin{equation}
\pb{\Pi_N,N}=-N \ , \pb{\Pi_N,\pi_N}=\pi_N \ , \pb{\Pi_N,a}=0
\end{equation}
and hence $\pb{\Pi_N,\mH_T}=0$. In other words we have second global
constraint $\Pi_T\approx 0$. We again split $\mC$ into $\infty-1$
local constraints $\tmC$ and one global constraint $\Pi_T\approx 0$
where we define $\tmC\approx 0$ as
\begin{equation}
\tmC=\mC-\frac{\gamma}{\int dx\gamma N}\Pi_N
\end{equation}
that obeys $\int dx N\tmC=0$. 
To proceed further we introduce  united notation for the
second class constraints as $\Psi_A=(\tpi_N,\tmC)$. Since clearly
$\pb{\tmC(x),\tmC(y)}\neq 0$ we find that the matrix of Poisson
brackets has schematic  form
\begin{equation}
\pb{\Psi_A(x),\Psi_B(y)}=\triangle_{AB} \equiv
\left(\begin{array}{cc} 0 & X \\
Y & M \\ \end{array}\right)
\end{equation}
so that the inverse matrix $\triangle^{AB}$ has the form
\begin{equation}
\triangle^{AB}=\left(\begin{array}{cc} -Y^{-1}M X^{-1} & Y^{-1} \\
X^{-1} & 0 \\ \end{array}\right) \ . 
\end{equation}
%
As the final step we have to ensure that $\Pi_T$ and $\Pi_N$ are the
first class constraints. $\Pi_N$ clearly is since it has vanishing
Poisson brackets with all constraints on the constraints surface. In
case of $\Pi_T$ this is not true but we can introduce following
combination of the constraints
\begin{equation}
\tPi_T=\Pi_T-\pb{\Pi_T,\Psi_A}\triangle^{AB}\Psi_B
\end{equation}
that obeys the equation
\begin{equation}
\pb{\tPi_T,\Psi_A}=\pb{\Pi_T,\Psi_A}-
\pb{\Pi_T,\Psi_C}\triangle^{CB}\pb{\Psi_B,\Psi_A} =0 \ , \quad 
\pb{\tPi_T,\tPi_T}=0 \ .
\end{equation}
For further purposes it is useful to determine explicit form of
$\tPi_T$. First of all we calculate Poisson bracket between $\Pi_T$
and $\tpi_N$ 
\begin{equation}
\pb{\Pi_T,\tpi_N(x)}=\pb{\Pi_T,\pi_N(x)}=-\pb{\int dy N\tmH_T,
\pi_N(x)}=\mC(x)\approx 0 \ .
\end{equation}
In case of the constraint $\mC$ we  only need to know that
this Poisson bracket is non-zero. Schematically we have
$\pb{\Pi_T,\vec{\Psi}^T}=(0,\ast)$, where $\ast$ is non-zero expression. 
Then we obtain 
\begin{equation}
\tPi_T=\Pi_T-(0,\ast)\left(\begin{array}{cc} -Y^{-1}M X^{-1} & Y^{-1} \\
X^{-1} & 0 \\ \end{array}\right)\left(\begin{array}{cc} 0 \\
\ast \\ \end{array}\right)=\Pi_T-\ast X^{-1}\tpi_N \
\end{equation}
which is very important result that shows that $\tPi_N$ does not
depend on the  constraint $\tmC\approx 0$.

Now we proceed to the analysis of the 
 constraint $\mH_1$. We add to it following
 expression proportional to the primary constraint  $\pi_N\approx 0$ 
\begin{equation}
\tmH_1=\mH_1+\pi_N\partial_1 N =-\gamma
\partial_1\pi+P\partial_1 B+
\pi_N\partial_1 N
\end{equation}
and introduce its smeared form
\begin{equation}
\bH_S(M^1)=\int dx M^1\tmH_1
\end{equation}
that has following Poisson brackets with canonical variables
\begin{eqnarray}
\pb{\bH_S(M^1),\gamma}&=&-(M^1\gamma)' \ , \nonumber \\
\pb{\bH_S(M^1),\pi}&=&-M^1\pi' \ , \nonumber \\
\pb{\bH_S(M^1),B}&=&-M^1 B' \ , \nonumber \\
\pb{\bH_S(M^1),P}&=&-(M^1P)'\ , \nonumber \\
\pb{\bH_S(M^1),N}&=&-M^1 N'\ , \nonumber \\
\pb{\bH_S(M^1),\pi_N}&=&-(M^1\pi_N)' \  \nonumber \\
\end{eqnarray}
and also
\begin{equation}
\pb{\bH_SM^1),a}=-(M^1)'a-M^1 a' \ .
\end{equation}
From these Poisson brackets we see that all constraints have
vanishing Poisson brackets with $\bH_S$ on constraint surface and
hence $\tmH_1\approx 0$ is the local first class constraint.

Let us now return to the second class constraints $\Psi_A$ and try 
to find their solutions. The problem is that these 
second class constraints contain the global first class constraints 
in their definition.  For that
reason it is natural to fix the global first class constraints by
appropriate global gauge fixing functions. Note that $\Pi_N$
generates pure time dependent rescaling of $N$ and $\pi_N$. For that
reason it is natural to introduce following gauge fixing function
\begin{equation}
\bG_N=\int dx \gamma N-C\approx 0 \ ,
\end{equation}
where $C$ is a constant
\footnote{In principle this could be time dependent function 
but we consider it to be constant for simplicity.}.
 Now this gauge fixing function has non-zero
Poisson bracket
\begin{equation}
\pb{\Pi_N,\bG_N}=-\int dx \gamma N=-C\neq 0 \ . 
\end{equation}
However this is not the end of the story due to the presence of the
second global constraint $\tPi_T$. We have to fix this first class
constraint in order to be able to solve $\mC\approx 0$ for $N$. Let
us propose following gauge fixing function
\begin{equation}
\bG_T=\int dx \gamma \pi-C_\pi(t)\approx 0 \ ,
\end{equation}
where we have to presume non-trivial dependence of $C_\pi$ on time
in order to find non-trivial dynamics. It is also easy to see that
\begin{equation}
\pb{\bH_S(N^1),\bG_T}=0 
\end{equation}
and also
\begin{eqnarray}
& &\pb{\bG_T,\tPi_T}\approx \pb{\bG_T,\Pi_T}=\nonumber \\
&=&\Pi_T-\int dx N(\frac{\kappa}{2\mu \gamma}\pi
P-\frac{2\gamma}{\kappa}U(B))\approx
-\int dx N\left(\frac{\kappa}{2\mu \gamma}\pi
P-\frac{2\gamma}{\kappa}U(B)\right)
 \ . \nonumber \\
\end{eqnarray}
Finally we  fix
the diffeomorphism constraint.
There is a number of possibilities
how to fix it. For example, we could use the gauge fixing condition
$\gamma=1$. However this condition does not fix the gauge completely
and there remains global diffeomorphism. For that reason we consider
another possibility when we impose the gauge fixing function
\begin{equation}
\mG_C=B-f(x) \ ,
\end{equation}
where $f(x)$ is prescribed function that obeys regularity condition
at infinity. Then we have
\begin{equation}
\pb{\tmH_1(x),\mG_C(y)}=B'(x)\delta(x-y)\approx f'(x)\delta(x-y) \ .
\end{equation}
Now we are ready to analyze the time evolution of all constraints and gauge
fixing functions in order to show that all Lagrange multipliers are fixed.
Recall that the total Hamiltonian with
gauge fixing functions included has the form
\begin{equation}
H_T=(1+\lambda_T) \tPi_T+\lambda_N \Pi_N+ V_T\bG_T+V_N\bG_N + \int dx(
\omega^A\Psi_A+ N^1\tmH_1+ M_1\mG_C) \ , 
\end{equation}
where we extended the  original Hamiltonian $\Pi_T$ in order to coincide
with $\tPi_T$ by appropriate linear combinations of constraints.

First of all we start with the constraint $\tmH_1\approx 0$. Since
the Hamiltonian was diffemorphism invariant we find
\begin{equation}
\partial_t \tmH_1(x)=\pb{\tmH_1(x),H_T}\approx 
\int dyM_1(y)\pb{\tmH_1(x),\mG_C(y)}=M_1 f'(x) \ . 
\end{equation}
Since by presumption  $f'(x)\neq 0$ for all $x$ we see that the only
possibility how to obey this equation is to demand that $M_1=0$.
Then the time evolution of the constraint $\tPi_T$ implies
\begin{equation}
\partial_t \tPi_T=\pb{\tPi_T,H_T}=V_T\pb{\tPi_T,\bG_T}=0
\end{equation}
which implies that $V_T=0$. In the same way time evolution of $\Pi_N$
implies
\begin{equation}
\partial_t \Pi_N=\pb{\Pi_N,H_T}=V_N\pb{\Pi_N,\mG_N}=0
\end{equation}
and we find $V_N=0$. However these results also imply that the time
evolution of the constraints $\Psi_A$ simplify considerably since
\begin{equation}
\partial_t \Psi_A(x)=\pb{\Psi_A(x),H_T}=\int
dy \omega^B(y)\pb{\Psi_A(x),\Psi_B(y)} =0
\end{equation}
due to the fact that  $V_T=V_N=M_1(x)=0$. Since the matrix of Poisson brackets of 
the second class constraints is non-singular we find that the equation above has the solution $\omega^B=0$.

 Finally we proceed to the requirement of the
preservation of the constraints $\bG_N,\bG_T$ and $\mG_C$.
In case of $\bG_T$ we
obtain
\begin{equation}
\partial_t \bG_T=\pb{\bG_T,H_T}=\partial_t
\bG_T+(1+\lambda_T)\pb{\bG_T,\tPi_T}=0
\end{equation}
using the fact that $\pb{\bG_T,\Pi_N}=0$. Then we obtain
\begin{equation}\label{lambdaTsol1}
\lambda_T=-1-\frac{\dot{C}_\pi}{\pb{\bG_T,\tPi_T}} \ .
\end{equation}
 In case of
$\bG_N$ we find
\begin{equation}
\frac{d\bG_N}{dt}=
\pb{\bG_N,H_T}=\lambda_N\pb{\mG_N,\Pi_N}
+(1+\lambda_T)\pb{\bG_N,\tPi_T}=0
\end{equation}
that can be solved form $\lambda_N$.
Finally the time evolution of the constraint $\mG_C$ has the form
\begin{eqnarray}
\partial_t \mG_C(x)&=&\pb{\mG_C(x),H_T}=
(1+\lambda_T )\pb{\mG_C(x),\tPi_T}+\int dy
N^1(y)\pb{\mG_C(x),\tmH_1(y)}=
\nonumber \\ 
&=&-(1+\lambda_T) N(\frac{\kappa}{2\mu^2\gamma}
B(1-\lambda)P+\frac{\kappa}{2\mu \gamma}\pi) +N^1(x)f'(x)=0 \ , 
 \nonumber \\
\end{eqnarray}
where we used the fact that $\tPi_T$ does not depend on $\tmC$. 
The previous equation can be solved 
 for $N^1$ as 
\begin{equation}\label{N1sol}
N^1=\frac{\kappa(1+\lambda_T)}{2\mu^2 \gamma f'}N(
B(1-\lambda)P+\pi) \ .  
\end{equation}
We see that we completely fixed all Lagrange multipliers.

 Now
we proceed to the analysis of the dynamics of the variables
$B,P,\pi,\gamma$ and $\pi_N$ and $N$. In case of $\pi_N$ 
we find that it is zero thanks to the constraint $\pi_N=0$.
$B$ is determined by the constraint $\mG_C=0$ that implies
\begin{equation}
B=f(x) \ . 
\end{equation}
Further, the conjugate momentum $P$ can be expressed using  the constraint
$\mH_1$ and we find 
\begin{equation}
P=\frac{\gamma \pi'}{f'(x)} \ .
\end{equation}

Finally we have to find $N$ as a function of dynamical variables $\pi,\gamma$. 
To do this we use the fact that  the constraint $\mC$ has the form
\begin{eqnarray}
\mC=
\frac{\kappa}{4\mu^2}B(1-\lambda)(\frac{\pi'}{B'})^2+\frac{\kappa}{2\mu}
\frac{\pi'}{B'} -\frac{1}{\kappa}U(B)-\frac{\beta}{
\kappa}\frac{B N'^2}{N^2}-\left(\frac{2\mu}{\kappa}B'\right)'
-\left(
\frac{2\beta}{\kappa}B \frac{N'}{N}\right)'=0 \ .\nonumber \\
\end{eqnarray}
Introducing variable $y=\frac{N'}{N}$ we can rewrite the equation
above to the form of the Riccati equation
\begin{equation}
y'=q_0(x)+q_1(x)y+q_2(x)y^2 \ ,
\end{equation}
where
\begin{eqnarray}\label{Nsol2}
q_0(x)&=&\frac{\kappa^2}{8\beta\mu^2}(1-\lambda) (\frac{\pi'}{B'})^2
+\frac{\kappa^2}{4\beta\mu}\frac{\pi'}{BB'}-\frac{1}{2\beta B}U(B)-
\frac{\mu}{\beta}(\frac{B'}{B})' \ , \nonumber \\
q_1(x)&=&-\frac{B'}{B} \ , q_2=-\frac{1}{2} \ . \nonumber \\
\end{eqnarray}
This equation can be explicitly solved as $N=N(\pi,\gamma)$ however
the explicit form of this solution is not important for us. 
We see that the remaining dynamical variables are $\pi,\gamma$ whose
equations of motion have the form
\begin{eqnarray}
& &\partial_t\pi(x)=\pb{\pi(x),H_T}=
\nonumber \\
& &=-(1+\lambda_T)\frac{N}{\gamma}(\frac{\kappa}{4\mu^2\gamma}B(1-\lambda)P^2
+\frac{\kappa}{2\mu\gamma}\pi P+\gamma
U(B)+\frac{2\mu}{\kappa}\frac{B'}{\gamma}a+\frac{\beta}{\kappa}\frac{Ba^2}{\gamma})
+\nonumber \\
& &\frac{\kappa\lambda_T}{2\mu^2 \gamma f'}N( B(1-\lambda)P+\pi)\pi '
\ , \nonumber \\
& &\partial_t \gamma(x)=\pb{\gamma(x),H_T}=
- N \frac{\kappa}{2\mu}P+\partial_1(N^1\gamma) \nonumber \\
\end{eqnarray}
using again the fact that $\tPi_T$ does not depend on
$\tmC$. It is important to stress that $N,N^1,\lambda_T$ all depend on 
$\gamma$ and $\pi$ as follow from (\ref{N1sol}) and (\ref{Nsol2}). Further, 
$1+\lambda_T$ is determined in (\ref{lambdaTsol1}) and we see that 
it is given as an integral over spatial section. In  summary, the equation
of motion for $\gamma,\pi$ are very complicated and is is not possible
to determine Hamiltonian on the reduced phase space. In other words, even 
$1+1 \ f(\tR)-$HL gravity has rather complicated structure so that it is hard to see whether it can be explicitly solved.

\section{The case $\lambda=1,\beta=0$}\label{fourth}
It is instructive to perform Hamiltonian analysis of the $f(\tR)-$HL gravity 
with special values of parameters.  In this section we consider the case 
 when $\lambda=1,\beta=0$ when 
the action has the form 
\begin{eqnarray}\label{actGR}
S=\frac{1}{\kappa}\int dt dx N\gamma (-2\mu
\nabla_n BK+2\frac{\mu}{\gamma^2}B'a-U(B)) \ . \nonumber \\
\end{eqnarray}
From  (\ref{actGR}) we obtain  conjugate momenta:
\begin{eqnarray}
\pi=-\frac{2\mu}{\kappa}\nabla_n B \ , \quad 
\pi_N\approx 0 \ ,  \quad \pi^1\approx 0 \ ,  \quad 
P=-\frac{2\mu}{\kappa}N\gamma K \nonumber \\
\end{eqnarray}
and hence the Hamiltonian has the form 
\begin{eqnarray}
H=\int dx (N\mH_T+N_1\frac{1}{\gamma^2}\mH_1) \ , \nonumber \\
\end{eqnarray}
where 
\begin{eqnarray}
\mH_T=
-\frac{\kappa}{2\mu }\pi P+\left(\frac{2\mu}{\kappa\gamma}B'\right)'+\frac{\gamma}{\kappa}U(B)
\ , \quad 
\mH_1=-\gamma\pi'
+P B' \ , \nonumber  \\
\end{eqnarray}
where we used integration by parts in order to have a theory linear in $N$. 
As usually the preservation of the primary constraints $\pi_N\approx 0 \ , 
\pi^1\approx 0$ implies the secondary constraints
 $\mH_T\approx 0 ,\mH_1\approx 0$. Now we have to analyze 
their preservation again. In order to do this we have to calculate corresponding
Poisson brackets of the smeared form of these constraints $\bH_T(X)=\int dx X\mH_T$
\begin{eqnarray}
\pb{\bH_T(X),\bH_T(Y)}=\int dx(XY'-X'Y)\frac{1}{\gamma^2}(PB'-\gamma \pi')=
\bH_S((XY'-YX')\frac{1}{\gamma^2}) \nonumber \\
\end{eqnarray}
and also 
\begin{eqnarray}
\pb{\bH_S(X^1),\bH_T(Y)}=\bH_T(-X^1Y') \ . \nonumber \\
\end{eqnarray}
We see that there is a crucial difference with the analysis performed in previous
sections since now there is local first class constraint $\mH_T\approx 0$ together
with spatial diffeomorphism constraint $\mH_1\approx 0$ and the first class constraints 
$\pi_N\approx 0 , \pi^1\approx 0$.

Let us now proceed to the gauge fixing of all constraints. At this place however we should be very careful with the variables $N$ and $N_1$. To see this in more details
remember that we are free to add secondary constraints $\mH_T,\mH_1$ with arbitrary 
Lagrange multipliers to the total Hamiltonian $H_T$. Let us also presume that
we couple the gravity with matter in the form of  free scalar field
\begin{equation}
S_{mat}=\frac{1}{2}\int dt dx N \gamma (\nabla_n\phi \nabla_n\phi-\frac{1}{\gamma^2}
\phi'^2)
\end{equation}
with corresponding matter contribution to the Hamiltonian in the form 
\begin{equation}
H_{matter}=\int dx [N(\frac{1}{2\gamma}P_\phi^2+\frac{1}{2\gamma}(\phi')^2)+N_1\frac{1}{\gamma^2}
P_\phi \phi'] \ . 
\end{equation}
Now when we include  the secondary constraints to the total 
Hamiltonian we find that it has the form 
\begin{equation}
H_{T,matter}=\int dx 
[(N+\lambda_T)(\frac{1}{2\gamma}P_\phi^2+\frac{1}{2\gamma}(\phi')^2)+(N_1+\lambda_1)\frac{1}{\gamma^2}
P_\phi \phi'] \ . 
\end{equation}
In order to return to the Lagrange formalism we have to calculate the equation of motion for $\phi$
\begin{equation}
\dot{\phi}=\pb{\phi,H}=(N+\lambda_T)P_\phi+(N_1+\lambda_1)\frac{1}{\gamma^2}\phi' \end{equation}
that allows us to express $P_\phi$ as 
\begin{equation}
P_\phi=\frac{1}{N+\lambda_T}(\dot{\phi}-(N_1+\lambda_1)\frac{1}{\gamma^2}\phi') \ .
\end{equation}
From this expression we immediately see that the components of the metric as 
it is seen by scalar field are  $N+\lambda_T, N_1+\lambda_1$ instead of the original ones. For that reason it is convenient to consider  $N,N_1$  as Lagrange multipliers and hence it  does not make 
sense to speak about their conjugate momenta and fix them. Rather we should fix $N,N_1$ by 
the requirement of the preservation of the gauge fixing functions  during the time evolution
\footnote{Alternatively, we can still keep $N$ and $N_1$ as dynamical fields and then it is possible to fix their values by fixing primary constraints $\pi_N\approx 0 \ , 
\pi^1\approx 0$. Then however gauge fixing of $\mH_T,\mH_1$ determine Lagrange multipliers $\lambda_T,\lambda_1$ that have to be included in the resulting metric as it is clear
from the discussion presented above.}. In other words 
the total Hamiltonian with gauge fixing constraints included has the form 
\begin{equation}
H_T=\int dx (N\mH_T+N^1\mH_1+\lambda^{\mH_T} \mG_{\mH_T}+\lambda^{\mH_1}\mG_{\mH_1}) \ .
\end{equation}
Of course, there is a freedom in the choice of the gauge fixing functions 
$\mG_{\mH_T},\mG_{\mH_1}$ when we only demand that  they have non-zero Poisson 
brackets with $\mH_T,\mH_1$. On the other hand when we impose the condition 
 that the solutions of the 
constraints correspond to the static solution we choose following form of these
constraints
\begin{equation}
\mG_{\mH_1}=\gamma^2-N\approx 0 \ , \quad \mG_{\mH_T}=P\approx 0  \ , 
\end{equation}
where now we have  following non-zero Poisson brackets
\begin{eqnarray}
& &\pb{\mG_{\mH_1}(x),\mH_T(y)}=-\frac{\kappa}{\mu}\gamma P
\approx 0 \ , \nonumber \\
& &\pb{\mG_{\mH_1}(x),\mH_1(y)}=	-\gamma(x)\gamma(y)\frac{\partial}{\partial y}\delta(x-y) \ , 
\nonumber \\
& &\pb{\mG_{\mH_T}(x),\mH_T(y)}=-\frac{2\mu}{\kappa}\frac{\partial}
{\partial y}\left(\frac{\partial_y \delta(x-y)}{\gamma}\right)-\frac{\gamma}{\kappa}
\frac{\delta U(B)}{\delta B}\delta(x-y) \ , \nonumber \\
& &\pb{\mG_{\mH_T}(x),\mH_1(y)}=-P\partial_y \delta(x-y)\approx 0 \ . \nonumber \\
\end{eqnarray}
Then the requirement of the preservation of the constraint $\mH_T\approx 0$ implies  \begin{eqnarray}
& &\partial_t \mH_T(x)=\pb{\mH_T(x),H_T}\approx 
\int dy \lambda^{\mH_T}(y)\pb{\mH_T(x),\mG_{\mH_T}(y)}=
\nonumber \\
&=&\frac{\gamma}{\kappa}\lambda^{\mH_T}\frac{\delta U}{\delta B}+
\frac{2\mu}{\kappa}\frac{\partial}{\partial x}
\left(\frac{\partial_x\lambda^{\mH_T}}{\gamma}\right) \nonumber \\
\end{eqnarray}
that has clearly solution $\lambda^{\mH_T}=0$. In the same way we find
\begin{eqnarray}
\partial_t\mH_1(x)=\pb{\mH_1(x),H_T}=2\gamma\partial_x(\lambda^{\mH_1}\gamma)=0
\end{eqnarray}
that has again solution $\lambda^{\mH_1}=0$. Let us proceed to the analysis
of the evolution of the constraints $\mG_{\mH_1}\approx 0,\mG_{\mH_T}\approx 0$
\begin{equation}
\partial_t \mG_{\mH_1}(x)=\pb{\mG_{\mH_1}(x),H_T}=
2\gamma(x)\partial_x(\gamma N^1)=0
\end{equation}
which is equal to zero for $N^1=\frac{D(t)}{\gamma}$ where $D(t)$ is 
arbitrary time dependent function. However in order to have a solution with 
the asymptotic behavior $N^1\rightarrow 0$ for $x\rightarrow \infty$ we
choose $D=0$. Then the requirement of the preservation of the constraint
$\mG_{\mH_T}$ has the form 
\begin{eqnarray}
& &\partial_t \mG_{\mH_T}(x)=\pb{\mG_{\mH_T}(x),H_T}=
\nonumber \\
&-&\frac{2\mu}{\kappa}\partial(\frac{\partial N}{N})-\frac{N^ 2}{\kappa}\frac{\delta U}{\delta B}=0 \ . 
\nonumber \\
\end{eqnarray}
It is convenient to parameterize $N$ as $N=e^\omega$ so that
the equation above has the form   
\begin{equation}
2\mu \omega''=-e^{2\omega}\frac{\delta U}{\delta B}
\end{equation}
that is generalization of the equation found in
\cite{Almheiri:2014cka} to the case
of $\mu\neq 1$.  Further, the 
Hamiltonian constraint on the constraint surface implies
\begin{equation}
2\mu \left(\frac{B'}{N}\right)'+NU(B)=0
\end{equation}
that can be written as
\begin{eqnarray}
2\mu B''-2\mu B'\omega'+e^{2\omega}U=0 \ .  \nonumber \\
\end{eqnarray}
This equation is again in agreement with the combinations 
of equations (2.14) and (2.15) presented in 
\cite{Almheiri:2014cka}.

\section{Non-projectable HL gravity with $f(x)=1$}\label{fifth}
Finally we perform  the Hamiltonian analysis of the special 
case when $f(x)=1$. 

To begin with note that in case $f(x)=1$ the equation of motion 
for $A$
implies that $B=1$ identically  and hence the action has the form
\begin{equation}
S=\frac{1}{\kappa}\int dt dx N\gamma \left((1-\lambda)K^2 -2\Lambda
+\beta a^2
\frac{1}{\gamma^2}\right) \ \nonumber \\
\end{equation}
which is the action studied in \cite{Li:2015itk}. However our goal is to carefully
identified global first class constraints so that we again proceed
to the Hamiltonian formulation of this theory.

Starting with the action (\ref{actHam}) we find following conjugate
momenta 
\begin{eqnarray}
\pi_N&=&\frac{\delta L}{\delta \dot{N}}\approx 0 \ , \quad 
\pi^1=\frac{\delta L}{\delta \dot{N}_1}\approx 0 \ , \nonumber \\
\pi&=&\frac{\delta L}{\delta
\dot{\gamma}}=\frac{2}{\kappa}(1-\lambda)K \ . \nonumber \\
\end{eqnarray}
Then it is easy to perform Legendre transformation in order to find
corresponding Hamiltonian
\begin{eqnarray}
H=\int dx(\pi\dot{\gamma}-\mL)=
\int dx (N\mH_T+N_1\frac{1}{\gamma^2}\mH_1) \ , \nonumber \\
\end{eqnarray}
where
\begin{eqnarray}
\mH_T=\gamma\left(\frac{\kappa}{4(1-\lambda)}\pi^2-\frac{\beta}{
\kappa}\frac{a^2}{\gamma^2}+\frac{2}{\kappa}
\gamma \Lambda\right)  \ ,  \quad
\mH_1=-\gamma\partial_1\pi \ . \nonumber \\
\end{eqnarray}
Again the requirement of the preservation of  the
primary constraints $\pi_N\approx 0 \ , \pi^1\approx 0$ implies 
 two secondary constraints 
\begin{eqnarray}
\partial_t\pi_N& &=\pb{\pi_N,H}=-\mH_T
-\frac{2\beta}{\kappa}\frac{1}{\gamma}a^2-\left(
\frac{2\beta}{\kappa}\frac{1}{\gamma}a\right)' =
\nonumber \\
&=&-\frac{\kappa}{4(1-\lambda)}\gamma \pi^2
-\frac{2}{\kappa}\gamma\Lambda -\frac{\beta}{\kappa}
\frac{a^2}{\gamma} -\left(
\frac{2\beta}{\kappa}\frac{1}{\gamma}a\right)' \equiv -\mC\approx
0 \  ,
 \nonumber \\
 \partial_t \pi^i&=&\pb{\pi^1,H}=-\mH_1\approx 0 \ , \nonumber \\
\end{eqnarray}
where  $\mC$ obeys the property
\begin{equation}
\int dx N \mC=\int dx N\mH_T \ .
\end{equation}
Now we should proceed completely as in section (\ref{third}) 
and we will find that the theory has identical structure of constraints.
For that reason we immediately skip to the analysis of the gauge
fixed theory. 
We again fix the constraint $\Pi_N$ with the gauge fixing function
\begin{equation}
\bG_N=\int dx \gamma N-C\approx 0 \ ,
\end{equation}
where $C$ is a constant. Now this gauge fixing function has non-zero
Poisson bracket
\begin{equation}
\pb{\Pi_N,\mG_N}=-\int dx \gamma N=-C\neq 0 \ ,
\end{equation}
while we have 
\begin{eqnarray}
\pb{\bG_N,\tpi_N(x)}=0 \ , \quad 
\pb{\bH_S(N^1),\bG_N}=0 \ . \nonumber \\
\end{eqnarray}
The  global constraint $\tPi_T$ is fixed by  gauge fixing function
\begin{equation}
\bG_T=\int dx \gamma \pi-C_\pi(t)\approx 0 \ 
\end{equation}
so that 
\begin{equation}
\pb{\bG_T,\Pi_T}=\int dx N\mH_T-\frac{4\beta}{\kappa^2} \int dx
N\gamma \Lambda\approx 
-\frac{4\beta}{\kappa}\Lambda C 
\end{equation}
and we see that $\bG_T$ cannot be gauge fixing function in case when
$\Lambda=0$ since in this case the theory possesses global
scale gauge symmetry with $\bG_T$ corresponding generator. We return 
to this problem below.  It
is also easy to see that
\begin{equation}
\pb{\bH_S(N^1),\bG_T}=0 \ . 
\end{equation}
Finally we fix the diffemorphism constraint using the 
gauge fixing condition 
\begin{equation}
\mG_S:\gamma-g(t)\approx 0 \ ,
\end{equation}
where $g(t)$ is an arbitrary time dependent function. 
Note that $\mG_S$ has following non-zero Poisson bracket with 
$\bH_S(N^1)$
\begin{equation}
\pb{\mG_S(x),\bH_S(N^1)}=(N^1)'
\end{equation}
that is zero for $N^1=N^1(t)$.  Now from $\mH_1=0$ we 
find that $\pi=\pi(t)$ and then 
the gauge fixing condition $\bG_T$ implies 
\begin{equation}
\int dx \gamma \pi(t)=g(t)\pi(t)\int dx=C_\pi(t) 
\end{equation}
and hence
\begin{equation}\label{pisol}
\pi(t)=\frac{C_\pi(t)}{g(t)L} \ ,
\end{equation}
where $L$ is regularized length of the system. 

Now we are ready to  determine  Lagrange multipliers for all
constraints and gauge fixing functions. Recall  that the total Hamiltonian with
gauge fixing functions included has the form
\begin{equation}
H_T=(1+\lambda_T) \tPi_T+\lambda_N \Pi_N+ V_T\bG_T+V_N\bG_N + \int dx(
\omega^A\Psi_A+ N^1\tmH_1+ M_1\mG_S) \ .
\end{equation}
From the previous expression we see that the effective lapse is $(1+\lambda_T)N$ instead
of $N$.  However
the value of $\lambda_T$ is fixed by the requirement of the preservation of all 
constraints. 
First of all we start with the constraint $\tmH_1\approx 0$. Since
the Hamiltonian is diffemorphism invariant we find
\begin{equation}
\partial_t \tmH_1(x)=\pb{\tmH_1(x),H_T}
=\int dyM_1(y)\pb{\tmH_1(x),\mG_C(y)}=M'_1(x)=0
\end{equation}
and this is equal to zero for $M^1=M^1(t)$. On the other hand we have to demand
that the Lagrange multipliers have correct asymptotic behavior at infinity 
so that  the only 
possible solution is $M^1=0$. 

Then the time evolution
of the constraint $\tPi_T$ implies
\begin{equation}
\partial_t
\tPi_T=\pb{\tPi_T,H_T}=V_T\pb{\tPi_T,\bG_T}=0
\end{equation}
which implies $V_T=0$. In the same way time evolution of $\Pi_N$
implies
\begin{equation}
\partial_t \Pi_N=\pb{\Pi_N,H_T}=V_N\pb{\Pi_N,\mG_N}=0
\end{equation}
and we find $V_N=0$. Then exactly as in section (\ref{third}) we 
find that $\omega^A=0$. 

 Finally we proceed to the requirement of the
preservation of the constraint $\bG_T,\bG_N$ and $\mG_S$. 
In case of $\bG_T$ we
obtain
\begin{equation}
\partial_t \bG_T=\pb{\mG_T,H_T}=\partial_t
\mG_T+(1+\lambda_T)\pb{\bG_T,\tPi_T}=0
\end{equation}
and hence we find
\begin{equation}\label{l1}
\lambda_T=-1-\frac{\dot{C}_\pi}{\pb{\bG_T,\tPi_T}}=-1-
\frac{\dot{C}_\pi}{\frac{4\beta}{\kappa}\Lambda C} \ .
\end{equation}
In case of
$\bG_N$ we find
\begin{equation}
\partial_t\bG_N=\pb{\bG_N,H_T}=\lambda_N\pb{\bG_N,\Pi_N}+(1+\lambda_T)\pb{\bG_N,\Pi_T}=0
\end{equation}
using the fact that $\omega^A=0$. The previous equation can be solved for $\lambda_N$
but the explicit solution is not important for us. 
Finally the time evolution of the constraint $\mG_S$ has the form
\begin{eqnarray}
& &\partial_t \mG_S(x)=\frac{\partial \mG_S}{\partial t}+\pb{\mG_S(x),H_T}
\nonumber \\
&-&\dot{g}+
(1+\lambda_T) \pb{\mG_S(x),\tPi_T}+\int dy
N^1(y)\pb{\mG_S(x),\tmH_1(y)}
\nonumber \\
&=&
-\dot{g}+(1+\lambda_T)\frac{\kappa}{2(1-\lambda)} N\gamma \pi+
N'_1 g(t)=0
 \ .  \nonumber \\
\end{eqnarray}
The previous equation can be solved for $N_1$ at least in principle. However
it is important to stress that $N_1$ is off diagonal component of the metric
so that if we demand that the metric is diagonal we have to impose the condition 
$N_1=0$. Then the previous equation implies
\begin{equation}\label{dotg1}
\dot{g}=(1+\lambda_T)\frac{\kappa}{2(1-\lambda)}Ng(t)\pi=
-\frac{\kappa}{8\beta}C_\pi \dot{C}_\pi N \ , 
\end{equation}
where in the final step we used (\ref{pisol}).
Let us now return to the condition $\mC=0$ that can be solved for 
$N$. However we simplify the calculation considerably when we
demand that the $(00)-$component of the  effective metric is equal to 
$-1$.  
This requirement implies that we have to demand 
that $N=\frac{1}{1+\lambda_T}$ that with the help of 
(\ref{l1}) implies 
\begin{equation}\label{Nsol1}
N=-\frac{\frac{4\beta}{\kappa}\Lambda C}{\dot{C}_\pi}
\end{equation}
and hence
\begin{equation}\label{dotg}
\dot{g}=-\frac{\kappa}{2(1-\lambda)}C_\pi \ . 
\end{equation}
Note that  (\ref{Nsol1}) implies that $N=N(t)$ and then the constraint 
$\mC$ simplifies considerably and leads to the result  
\begin{equation}
C^2_\pi=-\frac{8(1-\lambda)L^2}{\kappa^2}g^2\Lambda \ . 
\end{equation}
Inserting this expression into 
(\ref{dotg1}) we obtain a differential equation for $g$ 
\begin{equation}
\dot{g}=\pm\sqrt{\frac{2\Lambda L^2}{\lambda-1}}g
\end{equation}
that can be easily integrated with the result 
\begin{equation}
g=Ce^{\pm \sqrt{\frac{2\Lambda L^2}{\lambda-1}}t} \ . 
\end{equation}
In other words we found in the process of the gauge fixing
that all dynamical fields are fixed and that the line element 
has the form 
\begin{equation}
ds^2=-dt^2+Ce^{\pm \sqrt{\frac{2\Lambda L^2}{\lambda-1}}t} d^2x
\end{equation}
which is in complete agreement with the result derived in 
\cite{Li:2015itk}.

Finally we briefly mention the case of zero cosmological constant 
$\Lambda=0$. In this case we cannot use
the gauge fixing function $\bG_T=\int dx \gamma\pi$ since it
commutes with $\Pi_T$. Let us propose another gauge fixing function
\begin{equation}
\bG_T(f)=\int dx \gamma f(\pi)-C_\pi(t) \ ,
\end{equation}
where $\pb{\bH_S(N^1),\bG_T(f)}=0$ which follows from the fact that
$\pi$ is scalar. Using this gauge fixing function we find
\begin{eqnarray}
\pb{\bG_T(f),\Pi_T}=\int dx
(\frac{\kappa^2\gamma\pi}{4(1-\lambda)}(2f-\frac{df}{d\pi}\pi)-\frac{\beta^2}{
	\kappa^2}\frac{ a}{\gamma}\frac{df}{d\pi}) \nonumber \\
\end{eqnarray}
that is clearly non-zero and which also does not vanish on the
constraint surface. Then we can proceed as in previous case. First of all
the 
gauge fixing of the diffeomorphism constraint implies $\pi=\pi(t)$.
Further, if we demand that $(00)-$component of the effective metric
is equal to $-1$ we immediately obtain that $N=N(t)$ and hence 
$\pi=0$ as follows from $\mC=0$. If we again require that the metric
is diagonal we obtain equation (\ref{dotg1}) that implies $g=\mathrm{const}$ for $\pi=0$ and we can choose this constant to be equal to one. In other words, 
the flat line element 
\begin{equation}
ds^2=-dt^2+d^2x
\end{equation}
is the solution of the gauge fixed  $\Lambda=0$ non-projectable HL gravity 
 
To conclude,  we found
that in case of two dimensional non-projectable HL gravity 
all dynamical fields are fixed and there are no physical degrees of freedom left
which is in agreement with the analysis performed in 
\cite{Li:2015itk}.

 \noindent {\bf Acknowledgment}
\\
This work  was supported by the Grant Agency of the Czech Republic
under the grant P201/12/G028. 


\end{document}